\documentstyle[12pt]{article}
\newcommand{\beq}{\begin{eqnarray}}
\newcommand{\eeq}{\end{eqnarray}}
\newcommand{\la}{\langle}
\newcommand{\ra}{\rangle}

\begin{document}
\begin{flushright}
RYU-THP-93/1\\
January,\ 1993\\ 
\end{flushright}

\begin{center}

{\Large {\bf Nature of Chiral Transition in QCD }}\\ 
{\Large {\bf and Sigma Meson \ \ \ \ \ \ \ \ \ \ \ }}\\

\bigskip
\bigskip
\vspace{1cm}
{\large{Teiji Kunihiro\\

\bigskip
\bigskip

Faculty of Science and Technology, Ryukoku University,\\ 
Seta, Ohtsu-city, Japan}}\\
\end{center}

\vspace{1.5cm}
\begin{abstract}
After showing the characteristic features of the chiral transition 
 in QCD at 
 finite temperature $T$,
 we note the significance of the sigma-mesonic mode in the 
 chiral transition :  Sigma meson may be regarded as 
 the ``Higgs particle" in QCD, though the existence of such a mode in 
the real world is still unclear.  
We show the 
 properties of the sigma meson in hot and/or hadronic matter,
 from which
 we propose   experiments to reveal the existence the sigma meson clearly; 
 the experiments  include relativistic heavy-ion collisions 
to create hot hadronic matter, and electro-production in heavy nuclei.
\end{abstract}
\vspace{3cm}
{\large{\bf Talk Presented at Japan-China Joint Nuclear Physics
 Symposium ``Recent Topics on Nuclear Physics"}},\\
 {\bf November 30 - December 
 3, 1992,\ Tokyo Institute of Technology}\\

\newpage

\begin{center}

{\Large {\bf Nature of Chiral Transition in QCD }}\\ 
{\Large {\bf and Sigma Meson \ \ \ \ \ \ \ \ \ \ \ }}\\

\bigskip
\bigskip

Teiji Kunihiro\\

\bigskip

Faculty of Science and Technology, Ryukoku University,\\ 
Seta, Ohtsu-city, Japan\\
\end{center}

\begin{abstract}
We summarize the present status of the theoretical 
understanding of  the characteristic features of the chiral transition 
 in QCD at 
 finite temperature $T$.
 We emphasize  the significance of the sigma-mesonic mode in the 
 chiral transition :  Sigma meson may be regarded as 
 the ``Higgs particle" in QCD, though the existence of such a mode in 
the real world is still unclear.  
We show the 
 properties of the sigma meson in hot and/or hadronic matter,
 from which
 we propose   experiments to reveal the existence the sigma meson clearly; 
 the experiments  include relativistic heavy-ion collisions 
to create hot hadronic matter, and electro-production in heavy nuclei.
\end{abstract}

\section{Introduction}
The basic view-point which underlies the present report is  
that a change in the ground state (vacuum) as caused by that of
 the environment  may reflect in changes of properties of elementary excitations,
 hadrons in the case of QCD.  The salient features of QCD vacuum are
 (1) absence of free quarks and colored gluons, (2) dynamical breaking 
 of chiral symmetry, axial anomaly, approximate flavor-$SU(3)$ symmetry and so
 on.  In this report, we shall examine how , if any, properties of 
 Nambu-Goldstone bosons ($\pi, K, \eta$) and the scalar meson $\sigma$
 would change when the chiral symmetry is getting restored.
The present report is  based on Ref. \cite{character},
 done mostly in collaboration
 with T. Hatsuda.

\subsection{Significance of Sigma meson in 
 the chiral symmetry breaking in QCD}

What is  sigma meson? Sigma meson is
 iso-scalar and scalar meson with a low mass $\sim 600-700$ MeV.
 More specifically, sigma meson we refer to is the quantum fluctuation 
of   the order parameter $<<\bar qq>>$ of the chiral transition.  
 Thus in a sense, 
sigma meson is the Higgs particle in the chiral symmetry of QCD:
 In the standard
 electro-weak theory, the expectation value of a scalar field called 
Higgs
  field is the order parameter for the spontaneous breaking of the 
  $SU(2)\bigotimes U(1)$ symmetry, and the quantum fluctuation of the
   field in the new vacuum is the so called Higgs particle.
So we should seek sigma meson to demonstrate that the real world
 is realized due to the dynamical breaking of chiral symmetry,
 as eagerly as high-energy experimentalists search
 the Higgs particle. 
  
 Actually, the expected mass (the real part of the mass) of 
   sigma meson is about $600\sim 700$ MeV\cite{character,anomaly}, which could 
be seen in the    phase shift of the $\pi $-$\pi $ scattering in the $I=J=0$ channel.
However, the experimental phase shift does {\it not} show the expected resonance
 behaviour for the center-of-mass energy below 900 MeV; this is the main source
  of a  skeptical view about the existence of sigma meson.  
The resolution 
  of the skepticism is rather simple\cite{character}, though;
 the strong coupling of sigma with $2\pi $ gives rise a large imaginary 
mass or a large     width $\Gamma _{\sigma }\sim 450$ MeV of $\sigma $, 
because of which the energy
 where the phase shift cut 90 $^{\circ}$ will be shifted up to about 1 GeV where
  the experimental phase shift shows a resonance behavior with a so 
  complicated structure that the actual particle content contained 
in this   region are not well understood yet.
 Thus in the real world at $T=0$,
an experimental  demonstration of sigma meson might be very difficult.
Our point \cite{character} is, however, that once one can create 
 a high-temperature and/or high-density system by relativistic heavy-ion collisions, for example,
  one would have a good chance to see sigma meson as a {\it sharp}! 
resonance.

\subsection{Effective lagrangian approach}

 When one approaches these problems, some effective lagrangian or 
 model would be desirable in
  which the salient features of QCD are embodied. 
 Simulations of lattice QCD bear so strong constraints on 
  the computing ability that it is still awkward to compute various 
quantities
   freely, especially the dynamical aspects of the system.\cite{detar2}
   One of the merits of the studies based on effective theories lies in the 
fact that a physical idea or view can be easily put in  a calculation and 
thereby helps us getting physical insight into the problems under consideration:
  These will be helpful in future simulations on lattice QCD with forthcoming
   super computers.  Of course, it is also desirable if one can obtain
 a `BCS' theory for chiral symmetry breaking from QCD directly.
 
Our calculation is based on the generalized Nambu-Jona-Lasinio 
model 
\cite{njl,ms,tk,bjm} which embodies the explicit breaking of $SU_f(3)$ symmetry as given
  by  the flavor-dependent current quark masses, and the $U_{_A}(1)$ 
anomaly  given by the determinantal six-fermion interaction.

 The lagrangian we take is the following: 
\beq
{\cal L}_{NJL} & =&  \bar{q}(i \gamma \cdot \partial -{\bf m})q + \sum^{8}_{a=0}
{g_{_S} \over 2}[(\bar{q}\lambda_a q)^2 + (\bar{q}i\lambda_a
\gamma_5q)^2]\\ \nonumber
\ \ \ \  & \ \ &    +  g_{_D} [{\rm det}\bar{q}_i (1-\gamma_5) q_j +
h.c.],\\ \nonumber
& \equiv &{\cal L}_0 +{\cal L}_S + {\cal L}_{SB} + {\cal L}_D,\
\eeq
where the quark field $q_i$ has three colors ($N_c=3$) and three flavors
($N_f=3$), $\lambda_a$ ($a$=0$\sim$8) are the Gell-Mann matrices with
$\lambda_0$=$\sqrt{2 \over 3}\bf{1}$.
The second term is the explicit $SU_f(3)$-breaking part
with ${\bf m}={\rm diag}(m_u,m_d,m_s)$
being the current quark mass matrix.
The last term  is a reflection of the axial anomaly of
QCD,
which  has the $SU_{_L}(3) \otimes SU_{_R}(3)$-invariance but
breaks the  $U_A(1)$-symmetry.
This term  gives rise to   mixings of  the
different flavors both in the scalar and pseudo-scalar channels 
in the mean field approximation.

It is noteworthy that the NJL model can be cast into a form of a linear
 sigma model by integrating out the quark fields, and hence a non-linear
 sigma model with the parameters in the Lagrangian fixed by the 
 few parameters of  the underlying
 NJL model.  The relation among QCD, NJL model and sigma models for chiral
 transition has a
 good analogy with QED for electrons and ions, 
BCS theory, and phenomenological 
Gizburg-Landau model, as shown below:  
\begin{center}
\begin{tabular}{ccccc} 
QED & $\rightarrow$  & BCS & $\rightarrow$
 &  Ginzburg-Landau  model \\
 \ $\Updownarrow$ \ & ({\rm effective theory}) & \ $\Updownarrow$ \ &  ({\rm integrating out fermions})
 & \ $\Updownarrow$ \ \\
  QCD & $\rightarrow$ & NJL& $\rightarrow$  &  $\sigma$ models\\
\end{tabular}
\end{center}

\section{ Static Properties}
 
Recent lattice simulations show that the order and even the  existence 
 of the phase transition(s) are largely dependent on the number
 of the flavors especially when the physical current quark masses 
 are used\cite{ohsymp}: For $m_u\sim m_d\sim 10 {\rm MeV} << 100 {\rm MeV}
\stackrel {<}{\sim}m_s$, the phase transition may be weak 1st order
 or 2nd order or not exist. 

The gross feature of the $T$ dependence and the striking difference 
 between the condensates of u (d) quark and the s quark can
 be well described by the NJL model\cite{HK87}. It is noteworthy that at high
 temperatures, the flavor $SU(3)_f$-symmetry gets worse badly\cite{anomaly},
 which
 may reflect in the baryon and the vector meson spectra, because 
 they are well described by the constituent quark models.

The calculations based on the NJL model show \cite{HK87,anomaly} 
that the variation of the non-strange condensate with temperature is 
very large,
 while  that of the strange quark is moderate.  
This contrast between the non-strange and the strange sectors is
 also reflected in the change of the constituent quark masses.
 Hence one sees that
the $SU_f(3)$-symmetry is no more a good symmetry even approximately at
 temperatures larger than 150 MeV because the restoration of the chiral
 symmetry in the different sectors is achieved quite differently.  
One may also recognize that the approximate $SU_f(3)$-symmetry seen at 
$T=\mu _i=0$ is
rather accidentally realized by the spontaneous breaking of the chiral 
symmetry.\footnote{In discussing the
flavor symmetry in terms of the constituent quark masses, we are 
clearly taking not only the constituent quark picture of hadrons but
also  the  view that the constituent quark masses may be identified 
with the masses  generated by the spontaneous breaking of the chiral 
symmetry.}

Even away from the problem of the $SU_f(3)$-symmetry, the 
possible change of the spectra of  baryons and vector
mesons such as $\rho,\omega$ and $\phi $ mesons might provide us with 
a good signature of  the
formation of  hot hadronic matter: For instance, if one applies the
naive quark model to the vector mesons, our results tell us that 
these mesons would decrease
their masses as $T$ is raised, and the rate of the change are
larger in the non-strange vector mesons than in $\phi $ meson.
However the manifestation of the change might be more drastic for
$\phi $ meson because with a small change of the mass ($\sim 30$ MeV),
 $m_{\phi }$ gets into the subthreshold to the process of $\phi \rightarrow
 2K$.

\section{ Dynamic Properties --- Collective Excitations ---}

Lattice simulations\cite{detar2} and  effective theories 
\cite{character,mass}  predict the existence 
 of $\sigma $ meson, the mass of which decreases as $T$ is raised till
 $T_c$, a ``critical temperature": $m_{\pi}$ is found to be constant
 as long as $T<T_c$.\footnote{$T_c$ may be defined as the temperature
 at which $m_{\pi}$ starts to go high.}
It should be noted here that the lattice simulations only give the
 screening  masses, i.e., the mass-like parameters of the 
space-correlations
 of the hadronic composite operators, while the effective theories 
such as
 the NJL model can give real masses as well as screening ones.

The calculations with the NJL model \cite{character,anomaly,mass} show that pion hardly changes its 
mass but only starts become heavy near 
 $T\sim 200$ MeV, while sigma meson, which is found to be dominantly composed of
  the nonstrange sigma meson $\sigma _{_{NS}}\sim (\bar uu+\bar dd)/\sqrt 2$,
 decreases the mass $m_{\sigma }(T,\rho _B)$ as the chiral symmetry gets restored
and  eventually 
 $m_{\sigma }(T)$ becomes smaller than twice of pion mass $m_{\pi}$ 
at a temperature $T_{\sigma }\sim 190$ MeV.  This means that 
the width $\Gamma _{\sigma }$ of $\sigma $ meson due to the process 
$\sigma \rightarrow \pi \pi$ vanishes at $T_{\sigma }$.
It means that $\sigma $ meson  would appear as a sharp resonance at high temperature, 
though the large width $\Gamma _{\sigma }\sim 500$ MeV at $T=0$ 
prevent us from seeing $\sigma $ meson clearly at $T=0$ \cite{character}.  
Thus one sees that the results obtained in the two-flavor 
case \cite{character} are confirmed in the three-flavor case with 
the axial anomaly incorporated.  

   The behavior of kaon at finite 
 temperature was examined by the present author with use of the 
 $SU(3)$-NJL
 model including the anomaly term\cite{anomaly}.  It was found that as long as 
 the system is in the NG phase, the mass of kaon $m_{K}(T)$
 keeps almost a constant, the value   at $T=0$.

Then how about hadronic excitations at $T>T_c$. It is remarkable that 
 there seem exist colorless hadronic excitations even in the high-$T$
 phase\cite{mass} contrary to the naive picture of it.
There should exist
  precursory soft modes in the high temperature phase prior to 
the phase transition if the chiral transition is 
 of second order or weak first order:  The soft modes are actually fluctuations of the
 order parameter of the phase transition,
 $\la\la(\bar q q)^2\ra\ra$ and  hence 
 $\la\la(\bar q i\gamma _5\tau q)^2\ra\ra$
due to the chiral symmetry.
 We demonstrated these  using  an effective theory of QCD.\cite{mass}

  The lattice simulations \cite{lattice} showed that the screening masses of pion and sigma 
meson are both well below 
$2\pi T$, which indicates that the
interactions between q-$\bar {\rm q}$ in the pseudo-scalar and
 the scalar channels are still rather strong even in the high-$T$
 phase, as suggested in the NJL model. 

As for the correlations in the vector channel, see
\cite{qnum}.

\newpage

\section{ Implication to Experiment}

\begin{description}
\item[sigma meson I]
The sigma meson would decrease the mass while $m_{\pi}$
 keeps constant at high temperatures.  This suggests that at high temperatures
 the decay \ \ \
$\sigma \rightarrow 2\pi $\ \ \
 would get  suppressed and finally 
hindered, and them only the electro-magnetic process $\sigma \rightarrow
 2\gamma$ is allowable\cite{character,mass}. 
 It means that sigma meson may show up as a sharp
 resonance with the mass $m_{\sigma}\stackrel {>}{\sim} 2m_{\pi}$.  Thus
 we propose to observe $\pi ^{+}\pi^{-},\ 2\pi^0,\ 2\gamma$ and 
 construct the invariant mass and examine whether there is a bump 
 in the mass region 300 to 400 MeV.  
\item[sigma meson II]
 Recently, Weldon \cite{weldon} find that in the charged system, the 
 process \ \ \ 
$ \sigma \rightarrow \gamma \rightarrow \ {\rm 2 leptons}$\ \ \ 
 is possible, because $\pi^{+}$ and $\pi^{-}$ have  different
 chemical potentials, respectively.   
 The detection of lepton pairs would be hopeful because they
 interact with the matter only weakly in comparison with hadrons.
\end{description}

\section{Summary and concluding remarks}

We have discussed possible character change of several hadrons especially
 sigma meson.
 The  change is  associated  with the chiral restoration.
We have seen that   
 sigma meson would appear as a sharp resonance, decaying
 only by the electromagnetic process 
 $\sigma \rightarrow 
\gamma \gamma  $ with  an only tiny width and a low mass. 
   Therefore it would be interesting to detect $2\gamma '$s with 
   invariant masses of several hundred MeV in  relativistic heavy ion 
   collisions \cite{character}.

Recently there are some suggestions \cite{qcdsum} that
 the effects of chiral transition might be more significant at 
 finite baryonic density than at finite T.  At finite baryon density,
 there arises a vector-scalar coupling as is well-known in the
 $\sigma$-$\omega$ model\cite{weldon,qnum}. 
 On account of the coupling, it is possible 
 to create sigma meson in a nucleus by electron-nucleus scattering,
  due to the process $\gamma^{\ast} \  ({\rm virtual}) \rightarrow \sigma$.
  By measuring the decay products from $\sigma$ such as $2\pi (\pi ^{\pm}, 2\pi^0)$
 or $2\ell^{\pm}$, one would be able to see rather sharp resonance
 of the sigma meson.
  To make the experiment meaningful, one should examine the processes with
 a nucleon being emitted simultaneously for momentum matching. 
 Furthermore, to avoid the large 
 decay product from $\rho$ meson, the detection of neutral pions would 
 give clearer data for sigma meson.  Such experiment should be  feasible
 in CEBAF.

\bigskip

 \begin{center}
 {\large {\bf Acknowledgement}}
 \end{center}

The author thanks T. Hatsuda for collaboration for the 
 works on which the present report is based. 
 He is grateful to H. Shimizu for his interest in
 our work on sigma meson and discussions on experimental aspects of the
 problem.
The author  expresses his gratitude to the organizers of
 this symposium, especially to Professor M. Oka for giving 
him an opportunity to give a talk there.

\newpage

\newcommand{\btem}{\bibitem}
\newcommand{\TH}{T.\ Hatsuda}
\newcommand{\TK}{T.\ Kunihiro}
\newcommand{\PL}{Phys.\ Lett.\ {\bf B}}
\newcommand{\PTP}{Prog.\ Theor.\ Phys.}
\newcommand{\PR}{Phys.\ Rev.}
\newcommand{\PRL}{Phys.\ Rev. \ Lett.}
\newcommand{\NPB}{Nucl.\ Phys.\ {\bf B}}
\newcommand{\NPA}{Nucl.\ Phys.\ {\bf A}}

\end{document}